\documentclass[12pt]{article}
\usepackage{graphicx}
\usepackage{amsfonts}
 \def\fA{{\cal A}} \def\fB{{\cal B}}

\usepackage{theorem}
\theorembodyfont{\rmfamily}
\newtheorem{Lem}{Lemma}[section]
\newtheorem{Def}[Lem]{Definition}
\newtheorem{The}[Lem]{Theorem}
\newtheorem{Prop}[Lem]{Proposition}
\newtheorem{Cor}[Lem]{Corollary}

\newtheorem{Rem}[Lem]{Remark}

\newcommand{\qed}{\hbox{\rule{6pt}{6pt}}}

\begin{document}
\title{Fundamental properties of Tsallis relative entropy}
\author{S. Furuichi$^1$, K.Yanagi$^2$ and K.Kuriyama$^3$\\
$^1${\small Department of Electronics and Computer Science,}\\{\small Tokyo University of Science, Onoda City, Yamaguchi, 756-0884, Japan}\\{\small E-mail:furuichi@ed.yama.tus.ac.jp}\\
$^2${\small Department of Applied Science, Faculty of Engineering,} \\{\small Yamaguchi University,Tokiwadai 2-16-1, Ube City, 755-0811, Japan}\\{\small E-mail:yanagi@yamaguchi-u.ac.jp,}\\
$^3${\small Department of Applied Science, Faculty of Engineering,} \\{\small Yamaguchi University,Tokiwadai 2-16-1, Ube City, 755-0811, Japan}\\{\small E-mail:kuriyama@yamaguchi-u.ac.jp}}
\date{}
\maketitle
{\bf Abstract.} Fundamental properties for the Tsallis relative entropy in both classical and quantum systems are studied. 
As one of our main results, we give the parametric extension of the trace inequality
 between the quantum relative entropy and the minus of the trace of the relative operator entropy given by Hiai and Petz. 
The monotonicity of the quantum Tsallis relative entropy for the trace preserving completely positive linear map is 
also shown without the assumption that the density operators are invertible.
 The generalized Tsallis relative entropy is defined and its subadditivity is shown by its joint convexity. 
Moreover, the generalized Peierls-Bogoliubov inequality is also proven.
\vspace{3mm}

{\bf Keywords : } Tsallis relative entropy, relative operator entropy, monotonicity and generalized Peierls-Bogoliubov inequality 
\vspace{3mm}

\section{Introduction}
In the field of the statistical physics, Tsallis entropy was defined in \cite{Tsa} by 
$
S_q(X) = -\sum_{x} p(x)^q \ln_q p(x)
$
with one parameter $q$ as an extension of Shannon entropy, where $q$-logarithm is defined by $\ln_q(x) \equiv \frac{x^{1-q}-1}{1-q}$
for any nonnegative real number $q$ and $x$, and $p(x) \equiv p(X=x)$ is the probability distribution of the given randam variable $X$.
We easily find that the Tsallis entropy $S_q(X)$ converges to the Shannon entropy $-\sum_{x} p(x) \log p(x)$ as $q \to 1$, 
since $q$-logarithm uniformly converges to natural logarithm as $q \to 1$. 
Tsallis entropy plays an essential role in nonextensive statistics, which is often called Tsallis statistics, so that
many important results have been published from the various points of view \cite{AO}.
As a matter of course, the Tsallis entropy and its related topics are mainly studied in the field of statisitical physics.
However the concept of entropy is important not only in thermodynamical physics and statistical physics but also
in information theory and analytical mathematics such as operator theory and probability theory. 
Recently, information theory has been in a progress as quantum information theory \cite{NC}
with the help of the operator theory \cite{An,Furuta} and the quantum entropy theory \cite{OP}. To study a certain entropic quantity is much important for the
development of information theory and the mathematical interest itself. In particular, the relative entropy is fundamental in the sense 
that it produces the entropy and the mutual information as special cases. 
Therefore in the present paper, we study properties of the Tsallis relative entropy in both classical and quantum system.

In the rest of this section, we will review several fundamental properties of the Tsallis relative entropy, as giving short proofs for the convenience of the readers.
See \cite{BRT,Tsa2,Shi}, for the pioneering works of the Tsallis relative entropy and their applications in classical system.
\begin{Def}
We suppose $a_j$ and $b_j$ are probability distributions satsfying $a_j \geq 0, b_j \geq 0$ and $\sum_{j=1}^n a_j = \sum_{j=1}^n b_j =1$. Then we define
the Tsallis relative entropy between $A=\left\{a_j\right\}$ and $B=\left\{b_j\right\}$, for any $q \geq 0$ as
\begin{equation}
D_q(A\vert B) \equiv - \sum_{j=1}^n a_j \ln_q \frac{b_j}{a_j}
\end{equation}
where $q$-logarithm function is defined by $\ln_q (x) \equiv \frac{x^{1-q}-1}{1-q}$ for nonnegative real number $x$ and $q$, and we make a convention $0 \ln_q \infty \equiv 0$.
\end{Def}
Note that $\lim_{q\to 1} D_q (A\vert B) = D_1(A\vert B) \equiv \sum_{j=1} a_j \log \frac{a_j}{b_j}$, which is known as relative entropy (which is often called
Kullback-Leibler information, divergence or cross entropy).
For the Tsallis relative entropy, it is known the following proposition.
\begin{Prop}\label{prop1}
\begin{itemize}
\item[(1)] (Nonnegativity) $D_q(A\vert B) \geq 0$.
\item[(2)] (Symmetry) $D_q \left( {a_{\pi \left( 1 \right)} , \cdots ,a_{\pi \left( n \right)} \left| {b_{\pi \left( 1 \right)} , \cdots ,b_{\pi \left( n \right)} } \right.} \right) = D_q \left( {a_1 , \cdots ,a_n \left| {b_1 , \cdots ,b_n } \right.} \right).$
\item[(3)] (Possibility of extention) $D_q \left( {a_1 , \cdots ,a_n ,0\left| {b_1 , \cdots ,b_n ,0} \right.} \right) = D_q \left( {a_1 , \cdots ,a_n \left| {b_1 , \cdots ,b_n } \right.} \right).$
\item[(4)] (Pseudoadditivity) 
\begin{eqnarray*}D_q \left( {A^{\left( 1 \right)}  \times A^{\left( 2 \right)} \left| {B^{\left( 1 \right)}  \times B^{\left( 2 \right)} } \right.} \right) &=&
 D_q \left( {A^{\left( 1 \right)} \left| {B^{\left( 1 \right)} } \right.} \right) + D_q \left( {A^{\left( 2 \right)} \left| {B^{\left( 2 \right)} } \right.} \right) \\
&+& \left( {q - 1} \right)D_q \left( {A^{\left( 1 \right)} \left| {B^{\left( 1 \right)} } \right.} \right)D_q \left( {A^{\left( 2 \right)} \left| {B^{\left( 2 \right)} } \right.} \right),
\end{eqnarray*}
where $$A^{\left( 1 \right)}  \times A^{\left( 2 \right)}  = \left\{ {a_j^{\left( 1 \right)} a_j^{\left( 2 \right)} \left| {a_j^{\left( 1 \right)}  
\in A^{\left( 1 \right)} ,a_j^{\left( 2 \right)}  \in A^{\left( 2 \right)} } \right.} \right\},$$
$$B^{\left( 1 \right)}  \times B^{\left( 2 \right)}  = \left\{ {b_j^{\left( 1 \right)} b_j^{\left( 2 \right)} \left| {b_j^{\left( 1 \right)}  
\in B^{\left( 1 \right)} ,b_j^{\left( 2 \right)}  \in B^{\left( 2 \right)} } \right.} \right\}.$$
\item[(5)] (Joint convexity) For $0 \leq \lambda \leq 1$, any $q \geq 0$ and the probability distributions $A^{(i)}=\left\{a_j^{(i)}\right\}$,$B^{(i)}=\left\{b_j^{(i)}\right\}$, $(i=1,2)$,  we have 
\[
D_q \left( {\lambda A^{\left( 1 \right)}  + \left( {1 - \lambda } \right)A^{\left( 2 \right)} |\lambda B^{\left( 1 \right)}  + \left( {1 - \lambda } \right)B^{\left( 2 \right)} } \right) \le \lambda D_q \left( {A^{\left( 1 \right)} |B^{\left( 1 \right)} } \right) + \left( {1 - \lambda } \right)D_q \left( {A^{\left( 2 \right)} |B^{\left( 2 \right)} } \right).
\]
\item[(6)] (Strong additivity) 
\[
\begin{array}{l}
 D_q \left( {a_1 , \cdots ,a_{i - 1} ,a_{i_1 } ,a_{i_2 } ,a_{i + 1} , \cdots ,a_n \left| {b_1 , \cdots ,b_{i - 1} ,b_{i_1 } ,b_{i_2 } ,b_{i + 1} , \cdots ,b_n } \right.} \right) \\ 
  = D_q \left( {a_1 , \cdots ,a_n \left| {b_1 , \cdots ,b_n } \right.} \right) + b_i^{1 - q} a_i^q D_q \left( {\frac{{a_{i_1 } }}{{a_i }},\frac{{a_{i_2 } }}{{a_i }}\left| 
{\frac{{b_{i_1 } }}{{b_i }},\frac{{b_{i_2 } }}{{b_i }}} \right.} \right) \\ 
 \end{array}
\]
where $a_i  = a_{i_1 }  + a_{i_2 } ,b_i  = b_{i_1 }  + b_{i_2 }.$
\end{itemize}
{\bf (Proof)}
(1) follows from the convexity of the function $-\ln_q(x)$:
\[
D_q \left( {A|B} \right) \equiv  - \sum\limits_{j = 1}^n {a_j \ln _q \frac{{b_j }}{{a_j }}}  \ge  - \ln _q \left( {\sum\limits_{j = 1}^n {a_j \frac{{b_j }}{{a_j }}} } \right) = 0.
\]
(2) and (3) are trivial. (4) follows by the direct calculation. 
(5) follows from the generalized log-sum inequality \cite{BRT} :
\begin{equation}\label{logsum}
\sum\limits_{i = 1}^n {\alpha _i \ln _q \left( {\frac{{\beta _i }}{{\alpha _i }}} \right)}
  \le \left( {\sum\limits_{i = 1}^n {\alpha _i } } \right)\ln _q \left( {\frac{{\sum\limits_{i = 1}^n {\beta _i } }}{{\sum\limits_{i = 1}^n {\alpha _i } }}} \right),
\end{equation}
for nonnegative numbers $\alpha _i ,\beta _i \left( {i = 1,2, \cdots ,n} \right)$ and any $q \geq 0$.
We define the function $L_q$ for $q \geq 0$ to prove (6) as
\[
L_q \left( {x,y} \right) \equiv  - x\ln _q \frac{y}{x}
\]
and
\[
\left\{ \begin{array}{l}
 a_{i_1 }  = a_i \left( {1 - s} \right) \\ 
 a_{i_2 }  = a_i s \\ 
 \end{array} \right.,\,\,\left\{ \begin{array}{l}
 b_{i_1 }  = b_i \left( {1 - t} \right) \\ 
 b_{i_2 }  = b_i t .\\ 
 \end{array} \right.
\]
Then we have
\[
L_q \left( {x_1 x_2 ,y_1 y_2 } \right) = x_2 L_q \left( {x_1 ,y_1 } \right) + x_1 L_q \left( {x_2 ,y_2 } \right) + \left( {q - 1} \right)L_q \left( {x_1 ,y_1 } \right)L_q \left( {x_2 ,y_2 } \right),
\]
which implies the claim with easy calculations.
\hfill \qed
\end{Prop}

\begin{Rem}
\begin{itemize}
\item[1.] (1) of Proposition \ref{prop1} implies 
\[
S_q \left( A \right) \le  \ln _q n,
\]
since we have 
\[
D_q \left( {A|U} \right) =  - n^{q-1} \left( S_q \left( A \right) -   \ln_q n \right),
\]
for any $q \geq 0$ and two probability distributions $A=\left\{a_j\right\}$ and $U=\left\{u_j\right\}$, where  $u_j = \frac{1}{n}, (^{\forall}j)$,
where the Tsallis entropy is represented by
\[
S_q \left( A \right) \equiv  - \sum\limits_{j = 1}^n {a_j^q \ln _q a_j }. 
\]
\item[2.] (4) of Proposition \ref{prop1} is reduced to the pseudoadditivity for the Tsallis entropy:
\begin{equation}\label{cpseudo}
S_q(A^{(1)}\times A^{(2)}) = S_q(A^{(1)}) + S_q(A^{(2)}) +(1-q)S_q(A^{(1)}) S_q(A^{(2)}). 
\end{equation}
\item[3.] (5) of Proposition \ref{prop1} recover the concavity for the Tsallis entropy, by putting 
$B^{\left( 1 \right)}  = \left\{ {1,0, \cdots ,0} \right\}$,
$B^{\left( 2 \right)}  = \left\{ {1,0, \cdots ,0} \right\}.$
\item[4.] (6) of Proposition \ref{prop1} is reduced to the strong additivity for the Tsallis entropy:
\[
S_q \left( {a_1 , \cdots ,a_{i - 1} ,a_{i_1 } ,a_{i_2 } ,a_{i + 1} , \cdots ,a_n } \right) = S_q \left( {a_1 , \cdots ,a_{i - 1} ,a_i ,a_{i + 1} , \cdots ,a_n } \right) 
+ a_i ^q S_q \left( {\frac{{a_{i1} }}{{a_i }},\frac{{a_{i2} }}{{a_i }}} \right).
\]
\end{itemize}
\end{Rem}

We finally show the monotonicity for the Tsallis relative entropy. To this end, we introduce some notations.
We consider the transition probability matrix $W:\fA \to \fB$, which can be identified to the matrix having
the conditional probability $W_{ji}$ as elements, where $\fA$ and $\fB$ are alphabet sets (finite sets) and
 $\sum_{j=1}^m W_{ji} =1$ for all $i=1,\cdots ,n$.
By $A = \left\{ {a_i^{(in)} } \right\} $ and $B = \left\{ {b_i^{(in)} } \right\} $, 
two distinct probability distributions in the input system $\fA$ 
are denoted. 
Then the probability distributions in the output system $\fB$ are represented by
$WA = \left\{ {a_j^{(out)} } \right\},\,WB = \left\{ {b_j ^{(out)}} \right\}, $ where
$a_j^{(out)}  = \sum\limits_{i = 1}^n{ a_i^{(in)} W_{ji}} ,b_j^{(out)}  = \sum\limits_{i = 1}^n { b_i^{(in)} W_{ji}},$
in terms of $ W = \left\{ {W_{ji} } \right\}, \,\, (i = 1, \cdots ,n ; j = 1, \cdots ,m)$.
Then we have the following.
\begin{Prop}\label{prop3}
In the above notation, for any $q \geq 0$, we have
\[
D_q \left( {WA\left| {WB} \right.} \right) \le D_q \left( {A\left| B \right.} \right).
\]
{\bf (Proof)}
Applying the generalized log-sum inequality Eq.(\ref{logsum}), we have
\begin{eqnarray*}
 D_q \left( {WA\left| {WB} \right.} \right) &= & - \sum\limits_{j = 1}^m {a_j^{(out)} \ln _q \frac{{b_j^{(out)} }}{{a_j^{(out)} }}}  \\ 
&=&  - \sum\limits_{j = 1}^m {\sum\limits_{i = 1}^n {a_i^{(in)} W_{ji} } \ln _q \frac{{\sum\limits_{i = 1}^n {b_i^{(in)} W_{ji} } }}{{\sum\limits_{i = 1}^n {a_i^{(in)} W_{ji} } }}}  \\ 
&\leq &  -\sum\limits_{j = 1}^m  \sum\limits_{i = 1}^n {a_i^{(in)} {W_{ji} } \ln _q \frac{{ {b_i^{(in)} W_{ji} } }}{{ {a_i^{(in)} W_{ji} } }}}  \\ 
&=&  - \sum\limits_{i = 1}^n {a_i^{(in)} \ln _q \frac{{b_i^{(in)} }}{{a_i^{(in)} }}}   \\ 
&= & D_q \left( {A\left| B \right.} \right). \\ 
 \end{eqnarray*}
\hfill \qed
\end{Prop}

We note that the above proposition is a special case of the monotonicity of $f$-divergence \cite{Csi} for the convex function $f$.
As closing introduction, we should also note here that the Tsallis entropy can be derived by a simple transformation from R\'enyi entropy
which was used before Tsallis one in the mathematical literature. See \cite{AD} on the details of R\'enyi entropy, in particular see pp.184-191 of \cite{AD} for
the relation to the structural $a$-entropy \cite{HC} (or called the entropy of type $\beta$ \cite{Dar}) which is one of the nonextensive entropies including the Tsallis entropy.

\section{Quantum Tsallis relative entropy and its properties}

In references \cite{Ab2,Ab3}, the quantum Tsallis relative entropy was defined by
\begin{equation}\label{qqre}
D_q(\rho\vert \sigma) \equiv \frac{1-Tr[\rho^q\sigma^{1-q}]}{1-q} 
\end{equation}
for two density operators $\rho$ and $\sigma$ and $0 \leq q < 1$, as one parameter extension of the definition of the quantum relative entropy by Umegaki \cite{Um}
\begin{equation}
U(\rho\vert \sigma) \equiv Tr[\rho(\log \rho - \log \sigma)].
\end{equation}
See chapter II written by A.K.Rajagopal in \cite{AO}, for the quantum version of Tsallis entropies and thier applications.

For the quantum Tsallis relative entropy $D_q(\rho\vert \sigma)$ and the quantum relative entropy $U(\rho\vert \sigma)$, the following relations are known.

\begin{Prop}\label{opprop}(Ruskai-Stillinger \cite{RS} (see also \cite{OP}))
For the strictly positive operators with a unit trace $\rho$ and $\sigma$, we have,
\begin{itemize}
\item[(1)] $D_q(\rho\vert \sigma)\leq U(\rho\vert \sigma) \leq D_{2-q}(\rho\vert \sigma)$ for $0 \leq q<1$.
\item[(2)] $D_{2-q}(\rho\vert \sigma) \leq U(\rho\vert \sigma) \leq D_q(\rho\vert \sigma)$ for $1 < q \leq 2$.
\end{itemize}
\end{Prop}
Note that the both sides in the both inequalities converge to $U(\rho\vert \sigma)$ as $q \to 1$. 
We must extend the definition of the quantum Tsallis relative entropy Eq.(\ref{qqre}) for $0 \leq q \leq 2$ and impose the invertibility on the density operators of $D_{2-q}(\rho\vert \sigma)$ for $0 \leq q <1$ and of $D_q(\rho\vert \sigma)$ for $1 < q \leq 2$.

{\bf (Proof)}
Since we have for any $x >0$ and $t>0$,
$$ \frac{1-x^{-t}}{t} \leq \log x \leq \frac{x^t-1}{t},$$
the following inequalities hold for any $a,b,t>0,$
\begin{equation} \label{abt}
 a \left( \frac{1-a^{-t}b^t}{t}\right) \leq a \log\frac{a}{b} \leq a \left(\frac{a^tb^{-t}-1}{t}\right).
\end{equation}
Let $\rho=\sum_i \lambda_i P_i$ and $\sigma=\sum_j \mu_j Q_j$ be the spectral decompositions. 
Since $\sum\limits_i {P_i }  = \sum\limits_j {Q_j }  = I$, then we have 
 \[
\begin{array}{l}
 Tr\left[ {\frac{{\rho^{1 + t} \sigma^{ - t}  - \rho}}{t} - \rho\left( {\log \rho - \log \sigma} \right)} \right] \\ 
  = \sum\limits_{i,j} {Tr\left[ {P_i \left\{ {\frac{{\rho^{1 + t} \sigma^{ - t}  - \rho}}{t} - \rho\left( {\log \rho - \log \sigma} \right)} \right\}Q_j } \right]} \,\,\,\, \\ 
  = \sum\limits_{i,j} {Tr\left[ {P_i \left( {\frac{1}{t}\lambda _i^{1 + t} \mu _j^{ - t}  - \frac{1}{t}\lambda _i  - \lambda _i \log \lambda _i  + \lambda _i \log \mu _j } \right)Q_j } \right]}  \\ 
  = \sum\limits_{i,j} {\left( {\frac{1}{t}\lambda _i^{1 + t} \mu _j^{ - t}  - \frac{1}{t}\lambda _i  - \lambda _i \log \lambda _i  + \lambda _i \log \mu _j } \right)Tr\left[ {P_i Q_j } \right]}  \ge 0 .\\ 
 \end{array}
\]
Last inequality in the above is due to the inequality of the right side in Eq.(\ref{abt}).
Thus we have 
$$
Tr[\rho(\log \rho -\log \sigma)] \leq \frac{1}{t}Tr[\rho^{1+t}\sigma^{-t}-\rho].
$$
The left side inequality is proven by similar way.
Thus putting $1-q =t (>0)$ in the above, we have (1) in Proposition \ref{opprop}.
Also we have (2) in Proposition \ref{opprop}, by putting $q-1 = t (>0)$.

\hfill \qed

We next consider another relation on the quantum Tsallis relative entropy. In \cite{FuKa}, the relative operator entropy was defined by
$$
S(\rho\vert \sigma) \equiv \rho^{1/2}\log (\rho^{-1/2}\sigma\rho^{-1/2}) \rho^{1/2},
$$
for two strictly positive operators $\rho$ and $\sigma$. 
If $\rho$ and $\sigma$ are commutative, then we have $U(\rho\vert \sigma) = -Tr[S(\rho\vert \sigma)]$.
For this relative operator entropy and the quantum relative entropy $U(\rho\vert \sigma)$,
Hiai and Petz proved the following relation :
\begin{equation}\label{HP}
U(\rho\vert \sigma) \leq -Tr[S(\rho \vert \sigma)],
\end{equation}
in \cite{HP1} (see also \cite{HP2}).

In our previous papers \cite{YKF}, we introduced the Tsallis relative operator entropy $T_{q}(\rho \vert \sigma)$
as a parametric extension of the relative operator entropy $S(\rho\vert \sigma)$ such as
$$ 
T_{q}(\rho \vert \sigma) \equiv \frac{\rho^{1/2}(\rho^{-1/2}\sigma\rho^{-1/2})^{1-q}\rho^{1/2} -\rho}{1-q},
$$
for $0 \leq q < 1$ and strictly positive operators $\rho$ and $\sigma$, in the sense that 
\begin{equation}\label{lim1}
\lim_{q \to 1} T_{q}(\rho \vert \sigma) = S(\rho\vert \sigma).
\end{equation}
Actually we should note that there is a slightly difference between two parameters $q$ in the present paper and $\lambda$ in the previous paper \cite{YKF} which is an extension of \cite{gFuruta}.
If $\rho$ and $\sigma$ are commutative, then we have $D_q(\rho\vert \sigma) = -Tr[T_q(\rho\vert \sigma)]$.
Also we now have that 
\begin{equation}\label{lim2}
\lim_{q \to 1}D_{q}(\rho \vert \sigma) = U(\rho\vert \sigma).
\end{equation}
 These relations Eq.(\ref{HP}), Eq.(\ref{lim1}) and Eq.(\ref{lim2}) naturally lead us to show the following theorem as 
a parametric extension of Eq.(\ref{HP}).

\begin{The}\label{gHP}
For $0\leq q <1$ and any strictly positive operators with unit trace $\rho$ and $\sigma$, we have
\begin{equation}\label{eq:ghp}
D_{q}(\rho\vert \sigma) \leq -Tr[T_q(\rho\vert \sigma)]
\end{equation}
\end{The}
{\bf (Proof)}
We denote the $\alpha$-power mean $\sharp_{\alpha}$ by $A\sharp_{\alpha}B \equiv A^{1/2}(A^{-1/2}BA^{-1/2})^{\alpha}A^{1/2}$.
From Theorem 3.4 of \cite{HP2}, we have
$$
Tr[e^A \sharp_{\alpha}e^B] \leq Tr[e^{(1-\alpha)A+\alpha B}]
$$
for any $\alpha \in [0,1]$. Putting $A=\log\rho$ and $B=\log\sigma$, we have
$$
Tr[\rho\sharp_{\alpha}\sigma] \leq Tr[e^{\log\rho^{1-\alpha} +\log\sigma^{\alpha}}].
$$
Since the Golden-Thompson inequality $Tr[e^{A+B}] \leq Tr[e^Ae^B]$ holds for any Hermitian operators $A$ and $B$, we have
$$
Tr[e^{\log\rho^{1-\alpha} +\log\sigma^{\alpha}}] \leq Tr[e^{\log\rho^{1-\alpha}}e^{\log\sigma^{\alpha}}]=Tr[\rho^{1-\alpha}\sigma^{\alpha}].
$$
Therefore 
$$
Tr[\rho^{1/2}(\rho^{-1/2}\sigma\rho^{-1/2})^{\alpha}\rho^{1/2}] \leq Tr[\rho^{1-\alpha}\sigma^{\alpha}]
$$
which implies the theorem by taking $\alpha = 1- q$.

\hfill \qed

\begin{Cor}(Hiai-Petz \cite{HP1,HP2})
For any strictly positive operators with unit trace $\rho$ and $\sigma$, we have
\begin{equation}\label{oriHP}
Tr[\rho(\log\rho -\log \sigma)] \leq Tr[\rho \log(\rho^{1/2}\sigma^{-1}\rho^{1/2})].
\end{equation}
\end{Cor}
{\bf (Proof)}
It follows by taking the limit as $q \to 1$ in the both sides of Eq.(\ref{eq:ghp}).

\hfill \qed

Thus the result proved by Hiai and Petz in \cite{HP1,HP2} is recovered as a special case of Theorem \ref{gHP}.

For the quantum Tsallis relative entropy $D_q(\rho\vert \sigma)$,
(i) pseudoadditivity and (ii) nonnegativity are shown in \cite{Ab2}, moreover (iii) joint convexity and (iv) monotonicity for projective mesurements,
are shown in \cite{Ab3}.
Here we show the unitary invariance of $D_q(\rho\vert \sigma)$ and the monotonicity of that for the trace-preserving completely positive linear map.

\begin{Prop}
For $0 \leq q < 1$ and any density operators $\rho$ and $\sigma$, the quantum relative entropy $D_q(\rho \vert \sigma)$ has the following properties.
\begin{itemize}
\item[(1)] (Nonnegativity) $ D_q(\rho \vert \sigma) \geq 0 .$
\item[(2)] (Pseudoadditivity) $ D_q(\rho_1\otimes\rho_2 \vert \sigma_1 \otimes \sigma_2) = D_q(\rho_1\vert \sigma_1)+D_q(\rho_2\vert \sigma_2)
 +(q-1)D_q(\rho_1\vert \sigma_1)D_q(\rho_2\vert \sigma_2) .$
\item[(3)] (Joint convexity) $D_q(\sum_{j}\lambda_j\rho_j\vert \sum_j \lambda_j \sigma_j) \leq \sum_j \lambda_j D_q(\rho_j\vert \sigma_j). $
\item[(4)] The quantum Tsallis relative entropy is invariant under the unitary transformation $U$ :
$$D_q(U\rho U^*\vert U\sigma U^*) = D_q(\rho\vert \sigma). $$
\end{itemize}
\end{Prop}

{\bf (Proof)}
Since it holds $f(q,x,y)\equiv \frac{x-x^qy^{1-q}}{1-q} -(x-y) \geq 0$ for $x \geq 0, y \geq 0$ and $0\leq q <1$, we 
have $D_q(\rho \vert \sigma) \geq Tr[\rho - \sigma]$, which implies (1), since $\rho$ and $\sigma$ are density operators.
(See Proposition 3.16 of \cite{OP} on the so-called Klein inequality.) 

(2) follows by the direct calculation.

(3) follows from the Lieb's theorem that for any operator $Z$ and and $0\leq t \leq 1$, the functional $f(A,B) \equiv Tr[Z^*A^tZB^{1-t}]$ is joint concave with respect to
two positive operators $A$ and $B$.

(4) is obvious by the use of Stone-Weierstrass approximation theorem. (It also can be shown by the application of Theorem \ref{q_mono} in the below.)

\hfill \qed

(1) of the above proposition follows from the generalized Peierls-Bogoliubov inequality which will be shown in the next section.

In \cite{Petz}, the monotonicity for more generalized relative entropy was shown under the assumption of the invertibility of the density operators.
Here we show the monotonicity for the quantum Tsallis relative entropy in the case of $0 \leq q <1$ without the assumption of the invertibility of the density operators.

\begin{The}\label{q_mono}
 For any trace-preserving completely positive linear map $\Phi$, any density operators $\rho$ and $\sigma$ and $0\leq q <1$, we have
$$ D_q(\Phi(\rho) \vert \Phi(\sigma)) \leq D_q(\rho \vert \sigma). $$
\end{The}
{\bf (Proof)}
We prove this theorem as similar way in \cite{Lin}.
To this end, we firstly prove the monotonicity of $D_q(\rho\vert \sigma)$ for the partial trace $Tr_B$ in the composite sysytem $AB$.
Let $\rho^{AB}$ and $\sigma^{AB}$ be density operators in the composite system $AB$. 
From \cite{NC,Weh}, there exists unitary operators $U_j$ and the probability $p_j$ such that 
\[
\rho ^A  \otimes \frac{I}{n} = \sum\limits_j {p_j } \left( {I \otimes U_j } \right)  \rho ^{AB} \left( {I \otimes U_j } \right)^* ,
\]
where $n$ and $I$ present the dimension and identity operator of the system $B$,
 $\rho^A = Tr_B [\rho^{AB}]$ and $\sigma^A = Tr_B [\sigma^{AB}]$. 
By the help of the joint concavity and the unitary invariance of the Tsallis relative entropy, we thus have
\[
\begin{array}{l}
 D_q \left( {\rho ^A  \otimes \frac{I}{n}\left| {\sigma ^A } \right. \otimes \frac{I}{n}} \right) 
\le \sum\limits_j {p_j } D_q \left( { \left( {I \otimes U_j } \right)  \rho ^{AB}  \left( {I \otimes U_j } \right)  ^* \left| { \left( {I \otimes U_j } \right)  \sigma ^{AB}  \left( {I \otimes U_j } \right)  ^* } \right.} \right) \\ 
 \,\,\,\,\,\,\,\,\,\,\,\,\,\,\,\,\,\,\,\,\,\,\,\,\,\,\,\,\,\,\,\,\,\,\,\,\,\,\,\,\,\,\,\,\,\, 
= \sum\limits_j {p_j } D_q \left( {\rho ^{AB} \left| {\sigma ^{AB} } \right.} \right) \\ 
 \,\,\,\,\,\,\,\,\,\,\,\,\,\,\,\,\,\,\,\,\,\,\,\,\,\,\,\,\,\,\,\,\,\,\,\,\,\,\,\,\,\,\,\,\,\, 
= D_q \left( {\rho ^{AB} \left| {\sigma ^{AB} } \right.} \right). \\ 
 \end{array}
\]
Since $D_q \left( {\rho ^A  \otimes \frac{I}{n}\left| {\sigma ^A } \right. \otimes \frac{I}{n}} \right) = D_q \left( {\rho ^A \left| {\sigma ^A } \right.} \right)$, 
we thus have 
\begin{equation}\label{partial}
D_q(Tr_B(\rho^{AB}) \vert Tr_B(\sigma^{AB})) \leq D_q( \rho^{AB}\vert \sigma^{AB}) 
\end{equation}
It is known \cite{Sch} (see also \cite{Cho,Kra,Lin}) that every trace-preserving completely positive linear map $\Phi$ has 
the following representation with some unitary operator $U^{AB}$ on the total system $AB$ 
and the projection (pure state) $P^B$ on the subsystem $B$,
$$
\Phi(\rho^A) = Tr_B U^{AB}(\rho^A\otimes P^B) U^{AB^*}.
$$
Therefore we have the following result, by the result Eq.(\ref{partial}) and the unitary invariance of $D_q(\rho\vert \sigma)$ again,
\begin{eqnarray*}
D_q(\Phi(\rho^A) \vert \Phi(\sigma^A)) &\leq & D_q(U^{AB}  (\rho^A\otimes P^B)  U^{AB^*} \vert  U^{AB}(\sigma^A\otimes P^B)U^{AB^*} ) \\
& = & D_q(\rho^A\otimes P^B \vert \sigma^A\otimes P^B   ).
\end{eqnarray*}
which implies our claim, since $D_q(\rho^A\otimes P^B \vert \sigma^A\otimes P^B   ) =D_q(\rho^A \vert \sigma^A   ).$
\hfill \qed
\\

Putting $\sigma = \frac{1}{n}I$ in Theorem \ref{q_mono}, we have the following corollary.
\begin{Cor}
For any trace-preserving completely positive linear unital map $\Phi$, any density operator $\rho$ and $0\leq q <1$, we have
$$
H_q(\Phi (\rho)) \geq H_q(\rho),
$$
where $H_q(X) = \frac{Tr[X^q]-1}{1-q}$ represents the Tsallis entropy for density operator $X$, which is often called the quantum Tsallis entropy.
\end{Cor}

We note that Theorem \ref{q_mono} for the fixed $\sigma $, namely the monotonicity of the quantum Tsallis relative entropy in the case of $\Phi (\sigma) = \sigma$, was proved in \cite{Ab4} to establish Clausius' inequality.

\begin{Rem}
It is known \cite{Lin} (see also \cite{Ru}) that there is an equivalent relation between the monotonicity for the quantum
relative entropy and the strong subadditivity for the quantum entropy. However in our case, we have not yet found such a relation.  Because
the pseudoadditivity of $q$-logarithm function
$$\ln_qxy =\ln_qx +\ln_qy+(1-q)\ln_qx\ln_qy$$
disturbs us to derive the beautiful relation such as 
$$D_q(p(x,y)\vert p(x)p(y))= S_q(p(x)) + S_q(p(y)) -S_q(p(x,y))$$ 
for the Tsallis relative entropy $D_q(p(x,y)\vert p(x)p(y))$, the Tsallis entropy $S_q(p(x))$, $S_q(p(y))$ and the Tsallis joint entropy $S_q(p(x,y))$, even if our stage is in classical system.
\end{Rem}

\section{Generalized Tsallis relative entropy}

For any two positive operators $A$, $B$ and any real number $q \in [0,1)$, we can define the generalized Tsallis relative entropy.
\begin{Def}
$$D_q(A\vert \vert B) \equiv \frac{Tr[A]-Tr[A^qB^{1-q}]}{1-q}.$$
\end{Def}

To avoid the confusions of readers, we use the different symbol $D_q(\cdot \vert \vert \cdot)$ for the generalized Tsallis relative entropy.

Since Lieb's concavity theorem is available for any positive operators $A$ and $B$, the generalized Tsallis relative entropy has a joint convexity :
\begin{equation} \label{gjc}
D_q(\sum_{j} \lambda_j A_j \vert \vert\sum_{j} \lambda_j  B) \leq  \sum_j \lambda_j D_q(A_j \vert \vert B_j) ,
\end{equation}
for the positive number $\lambda_j$ satisfying $\sum_{j}\lambda_j =1$ and any positive operators $A_j$ and $B_j$.
Then we have the subadditivity of the generalized Tsallis relative entropy between $A_1+A_2$ and $B_1+B_2$.

\begin{The}
For any positive operaors $A_1,A_2,B_1$ and $B_2$, and $0 \leq q < 1$, we have the subadditivity
\begin{equation} \label{gthe}
D_q(A_1+A_2 \vert \vert B_1+B_2) \leq D_q(A_1 \vert \vert B_1) + D_q (A_2 \vert \vert B_2).
\end{equation}
\end{The}
{\bf (Proof)}
Firstly we note that we have the following relation for any numbers $\alpha$ and $\beta$, and two positive operators
$A$ and $B$,
\begin{equation} \label{first}
D_q(\alpha A \vert \vert \beta B) = \alpha D_q(A \vert \vert B) -\alpha \ln_q \frac{\beta}{\alpha} Tr[A^qB^{1-q}].
\end{equation}
Now from Eq.(\ref{gjc}), we have 
$$
D_q(  \lambda_1 X_1 + \lambda_2 X_2 \vert \vert  \lambda_1 Y_1 + \lambda_2 Y_2  ) \leq \lambda_1 D_q(X_1 \vert \vert Y_1) + \lambda_2 D_q(X_2 \vert \vert Y_2)    
$$
for any positive operators $X_1,X_2,Y_1$ and $Y_2$, and $\lambda_1$, $\lambda_2\,\, (\lambda_1 +\lambda_2=1)$.
Putting $A_i = \lambda_i X_i$ and $B_i = \lambda_i Y_i$ for $i=1,2$ in the above inequality, we have
$$
D_q(  A_1 + A_2 \vert \vert  B_1 + B_2  ) \leq \lambda_1 D_q( \frac{A_1}{\lambda_1} \vert \vert \frac{B_1}{\lambda_1} ) + \lambda_2 D_q(\frac{A_2}{\lambda_2} \vert \vert \frac{B_2}{\lambda_2} )    
$$
Thus we have the our claim due to Eq.(\ref{first}). 

\hfill \qed

%
%
%
%
As a famous inequality in statistical physics, the Peierls-Bogoliubov inequality \cite{Hua,BPL} is known.
Finally, we prove the generalized Peierls-Bogoliubov inequality for the generalized Tsallis relative entropy in the following.

\begin{The}\label{bogo}
For any positive operators $A$ and $B$, $0 \leq q < 1$,
$$D_q(A\vert \vert B) \geq \frac{Tr[A]-(Tr[A])^q(Tr[B])^{1-q}}{1-q}.$$
\end{The}

{\bf (Proof)}
In general, it holds the following H\"older's inequality
\begin{equation}\label{hol}
\vert Tr[XY]\vert \leq Tr[\vert X \vert ^s]^{1/s} Tr[\vert  Y\vert ^t ]^{1/t}
\end{equation}
for any bounded linear operators $X$ and $Y$ satisfying $Tr[\vert X\vert^s] < \infty $ and $Tr[\vert Y\vert^t] < \infty $ and for any $1<s<\infty$ and $1<t<\infty$ satisfying $\frac{1}{s} +\frac{1}{t}=1$.
By putting $X=A^q, Y=B^{1-q}$ and $s=\frac{1}{q}, t=\frac{1}{1-q}$ in Eq.(\ref{hol}), we have
$$ Tr[A^qB^{1-q}] \leq (Tr[A])^q (Tr[B])^{1-q},$$
which implies our claim.
\hfill \qed

Note that Theorem \ref{bogo} can be considered a noncommutative version of Eq.(\ref{logsum}).
If $A$ and $B$ are density operators, then the nonnegativity of the quantum Tsallis relative entropy follows from Theorem \ref{bogo}.

\section{Conclusion}
As we have seen, the monotonicity of the quantum Tsallis relative entropy for the trace-preserving
completely positive map was shown. Also the trace inequality between the Tsallis quantum relative entropy 
and the Tsallis relative operator entropy was shown. It is remakable that our inequality recovers the famous inequality shown by Hiai-Petz 
as $q \to 1$. 
\section*{Acknowledgement}
The authors thank referees for valuable comments to improve the manuscript.

\end{document}